\documentstyle[twocolumn,epsfig,aps]{revtex}
\newcommand{\ptl}{\partial}
  
\begin{document}
  \title {Laminar plasma dynamos}
  \draft
  \input{epsf}
  \author{Zhehui Wang \footnote{email: zwang@lanl.gov}, Vladimir I.~Pariev \footnote{also at University of Rochester, and at Lebedev Physical Institute, Leninsky Prospect 53, Moscow B-333,
117924, Russia }, Cris W.~Barnes, and Daniel C.~Barnes}
\address{Los Alamos National Laboratory, Los Alamos, NM 87545}

  \date{\today ,\hspace{0.1cm} accepted by Phys. Plasmas}
  \maketitle

  \begin{abstract}
A new kind of dynamo utilizing flowing laboratory plasmas has
    been identified. Conversion of plasma kinetic energy to
magnetic energy is verified numerically by kinematic
dynamo simulations for
magnetic Reynolds numbers above 210. As opposed to
intrinsically-turbulent liquid-sodium dynamos, The proposed plasma dynamos
correspond to laminar flow topology. 
Modest plasma parameters, 1--20~eV temperatures,
$10^{19}$--$10^{20}$~m$^{-3}$ densities in 0.3--1.0~m scale-lengths driven
by velocities on the order of the Alfv\'en Critical Ionization Velocity (CIV),
self-consistently satisfy the conditions needed for the magnetic field amplication.
Growth rates for the
plasma dynamos are obtained numerically with different geometry and magnetic Reynolds
numbers. Magnetic-field-free coaxial plasma guns can be used
to sustain the plasma flow and the dynamo.

  \end{abstract}


% LaTeX\ Article Template - using defaults

Dynamo action\cite{Moffat:1978,Kulsrud:1999} is believed to be the fundamental
mechanism that creates the magnetic field commonly observed from
planetary to galactic scales\cite{Parker:1992}.  Conversion of
kinetic (flow) energy into magnetic energy is the best known
explanation for observed magnetic fields in the universe that are
persistent against resistive diffusion.  
A direct experimental proof of the conversion of plasma flow
energy to magnetic energy has not been accomplished to date.
Another motivation of this letter is to show dynamo experiments
can be performed in plasmas, as an alternative to liquid
sodium or other conducting liquids\cite{Tilgner:2000}. The plasma dynamos discussed here are funadmentally different from what has generally been known for
spheromaks\cite{Jarboe:1994,Barnes:1986} and reversed field
pinches\cite{Schoenberg:1984,Ji:1996}, when the term `dynamo' has traditionally been invoked to
explain conversion of one type of magnetic flux into another (toroidal flux
into poloidal flux in the spheromak case). Because in these previous cases, the total magnetic field energy does not increase.  In the present case, the plasma flow energy is converted into magnetic energy, and magnetic field is amplified accordingly.

Existing and proposed laboratory dynamo experiments
\cite{Gailitis:2001,Reighard:2001,Peffley:2000} have used
liquid sodium as conducting medium. Liquid sodium has low resistivity and viscosity\cite{Roberts:1992} for
conduction of electricity and fluid flow, as well as a low melting
temperature which eases the creation of the liquid
state for the conductor. A constraint on liquid sodium dynamos is
intrinsically turbulent flow due to the small ratio of viscosity
to resistivity\cite{Roberts:1992}.  The turbulent flow makes it
difficult to make detailed comparisons between experiments and
theories. Also, resistivity of the liquid sodium is only variable
within a factor of two\cite{Reighard:2001}, which implies the
magnetic Reynolds number, $Re_m$, may be varied only within a
small range, primarily by variation of the experimental dimensions.  Even
with the high conductivity of the liquid sodium, existing technologies can
only move liquid sodium at velocities no more than 20~m/sec\cite{Roberts:1992}. Therefore
$\sim$1~meter in size is necessary to have a dynamo excitation for a 
liquid sodium experiment, and $Re_m$ has been limited to values
below a few hundred\cite{Reighard:2001,Roberts:1992}.

Dynamo action is described by the induction equation\cite{Moffat:1978,Kulsrud:1999,Roberts:1992}
\begin{equation}
\frac{\partial {\bf B}}{\partial t} = \nabla\times\left(\bf U
\times {\bf B}\right)+\frac{\eta}{\mu_0}\nabla^2 {\bf B},
\label{induction:1}
\end{equation}
where ${\bf U}$ is the fluid flow velocity, and $\eta$ is the
resistivity.  Without
the flow {\bf U}, the induction equation is a diffusion equation for
magnetic field ${\bf B}$ with the characteristic diffusion time
determined by the dimension of the field and resistivity. Only when
{\bf U} does not vanish may ${\bf B}$ grow by transfering the flow
energy. The magnetic Reynolds number $Re_m \equiv \mu_0 L U_0 / \eta$
measures the relative amplitude of the flow drive to diffusion in
Eq.~(\ref{induction:1}), where $L$, $U_0$ and $\eta$ are characteristic
values for the dimension, velocity field, and resistivity.
Theories\cite{Roberts:1992,Glatzmaier:1998} predict that only
when the magnetic Reynolds number $Re_m$ exceeds certain threshold
values (from 20 to 200, depending on the geometry), is dynamo action
possible.

$Re_m$ greater than 200 appears easily achievable in flowing laboratory
plasmas by this analysis.  Thus an experiment is possible to create magnetic field energy
that grows by transformation of the plasma flow energy alone.  Such a new plasma
dynamo can cover a wide range of magnetic Reynolds numbers both below and
above the threshold by controlling the plasma flow velocity and the plasma
resistivity.  In the entire plasma system, one may encounter various length
scales $L$, velocities $U$, as well as uncertain resistivity (the plasma
resistivity is often "anomalous").  Average, ``global'' values
characteristic of the entire flow appear reasonable to use for comparison
to theoretical threshold values of $Re_m$.  In the plasma regime considered
here where the flow is laminar (see below), the scale length $L$ is close
to the dimension of the boundary.  The characteristic velocity $U_0$ is
representative of a Beltrami-flow profile\cite{Kageyama:1999}, which is a flow satisfying
$\nabla\times{\bf U}=\lambda_B {\bf U}$ with constant parameter $\lambda_B$.
Spitzer resistivity provides a realistic 
estimate since the magnetic field is weak during the initial growth of the
magnetic field, and the long plasma lifetimes imply local thermal
equilibrium. In addition, energy balance is dominated by atomic physics
(radiation by the electrons and boundary losses by the ions) and classical
thermal conduction\cite{Braginskii:1965}, and the plasma temperatures will be close to being in
equilibrium and in the 1--20 eV range, even with substantial high-$Z$ impurities.

One key feature of these laboratory plasma dynamos is the 
high speed plasma flows (see Fig.~1) which will be on the order of the critical ionization
velocity (CIV)\cite{Alfven:1956}. CIV, denoted as $U_c$, is
defined as $U_c \equiv \sqrt{2 E_i/M_i}$, where $E_i$ is the
ionization energy and  $M_i$ the ion mass. A velocity
of this magnitude is routinely achieved using plasma accelerators
or thrusters\cite{Schoenberg:1998}. For hydrogen gas, $U_c$ is 51
km/sec.  For ease of estimation, the electron temperature determining the
resistivity can be assumed to be in equilibrium with the ion temperature
which is itself set to the ``flow equivalent temperature'' $T_i =
(M_iU^2)/2$.  With this simplifying assumption,  $Re_m$ scales as
$Re_m = 15 T_i^2$[eV] $L$ [m]. For $L$ about 0.5~m and $T_i$ ranging from 1
to 20~eV, the $Re_m$ varies from 7.5 to 
3000. The threshold $Re_m = 200$ corresponds to $T_i = 5.2$~eV, or a flow
velocity of 31 km/sec for hydrogen, and hence in the expected range of CIV
velocities.  Because of the 
high plasma flow velocity, $Re_m$ in a plasma dynamo can be well above the
threshold magnetic Reynolds number for magnetic field to grow, even
considering that the resistivity of the plasma is much greater
than that of liquid sodium. In Fig.~1, these plasma dynamos are
compared with existing and proposed sodium dynamos, which are the only
known laboratory dynamos so far.

Another unique feature of the new plasma dynamo is laminar flow. Flow
topology is determined by the kinetic Reynolds number, $Re$, which
characterizes the relative amplitude of the fluid convection to viscous
dissipation in the Navier-Stokes equation. The critical Reynolds number for
onset of turbulence $Re$ is more than a few thousand, and fluid motion is
laminar for $Re$ below this threshold\cite{Vasilev:1999}. Plasmas discussed
here can be approximated as fluids because the mean free path for particle
motion is much less than the system dimension. The kinetic Reynolds number $Re$ then scales with the device dimension
and plasma ion temperature as $Re=4.0\times 10^{-16} L$ [m]
$n_e$ [m$^{-3}$]/$T_i^2$ [eV].  An experimental dimension $L$ of 0.5~m is adequate to meet
the conditions of a plasma dynamo, with density 10$^{19}$ to
10$^{20}$~m$^{-3}$ and temperature 1--20~eV.  This corresponds to the
kinetic Reynolds number $Re$ varying from 5 to 20000, which imply that plasma flow will be mostly laminar.

Plasma flow can be created using coaxial plasma guns, or Marshall guns\cite{Jarboe:1994,Wang:2001}.  For
the gun plasma momentum to be effectively transferred to the bulk plasma in a chamber, the
mean free-path, $\lambda_{mfp}$, of the gun plasma ions should be
less than the size of the dynamo chamber, that is, $\lambda_{mfp} < L$ is
required.  This is equivalent to $Re > 2.5$, and hence the kinetic Reynolds
number cannot be too small.  The magnetic Prandtl number ($Pr_m$) is
defined as the ratio $Pr_m=Re_m /Re$ and scales as $T_i^4/n_e$ with values
in the range of 3.7$\times 10^{-4}$ to 600 for the plasma parameters
above. In comparison, the $Pr_m$ for liquid sodium experiments is $\sim
10^{-5}$. The plasma Prandtl number can be very similar to that of the
interior of the sun or the galactic plasma dynamos, albeit at very much
lower Reynolds numbers.

The gun-plasma momentum transfer is also affected by ion-neutral
collisions, which include ion-neutral momentum transfer and ion-neutral
charge exchange.  The ion-neutral momentum transfer cross section is 100
times less than that of ion-ion collision\cite{Goldston:1995}. Therefore, as long as the ionization
fraction is greater than 1\%, ion-neutral momentum transfer can be
neglected. The charge exchange cross section is about half of that of
ion-ion momentum transfer. Therefore ion-ion momentum transfer is the
dominant process for gun-plasma momentum transfer. Generation of toroidally
rotating plasmas is most efficient when the ionization fraction is greater
than 30\%.

A possible experimental configurationtion is shown in Fig.~2, when one `main gun' (Labeled as MG) is used together with three `toroidal guns' (TG, not shown) to produce toroidal and poloidal plasma flows. Kinetic energy density for a 13 eV and 10$^{19}$ m$^{-3}$ plasma is equivalent to a magnetic field strength of 53 gauss at the same energy density. Theoretical studies indicate that the fraction of the plasma flow kinetic energy that can be converted into magnetic field depends on the ratio of the magnetic Reynolds number to the kinetic Reynolds number\cite{Kleva:1995}. As much as 20\% of the flow energy was seen to convert into magnetic field numerically. If this is the case, magnetic field of up to 24 gauss may be generated when the back-reaction of the magnetic field on plasma flow is taken into account.

Numerical results confirm the existence of laminar plasma dynamos. The plasma flow is approximated by a Beltrami flow, $\nabla\times
{\bf U}=\lambda_B {\bf U}$, which corresponds to a state of
maximum kinetic helicity [$\int dV ({\bf U}\cdot {\nabla \times {\bf U}})$]
for a given total kinetic energy and a given boundary condition.  We
can use the Beltrami flow approximation because the flow is laminar here, large coherent flow pattern can be
established. The Beltrami flow is also an eigenmode for the cylindrical
boundary condition considered.  Theoretical results indicate that large kinetic helicity
content can lead to large dynamo growth rate. The solution of axisymmetric
Beltrami flow inside the cylinder $0<r<R$ and $-L/2<z<L/2$
%is
%\begin{eqnarray}
%&& v_r=J_1\left(j_{11}\frac{r}{R}\right)\frac{\pi}{L}\sin\frac{\pi
%z}{L}
%\mbox{,}\nonumber\\
%&&
%v_z=\frac{j_{11}}{R}J_0\left(j_{11}\frac{r}{R}\right)\cos\frac{\pi
%z}
%{L} \mbox{,}\label{eqn2}\\
%&& v_{\phi}=\lambda_B
%J_1\left(j_{11}\frac{r}{R}\right)\cos\frac{\pi z}{L}
%\mbox{,}\nonumber
%\end{eqnarray}
%where $J_0(x)$ and $J_1(x)$are
%the Bessel functions, $j_{11}$ is !the first root of !$J_1(x)=0$,
%$\lambda_B^2=\frac{j_{11}^2}{R^2}+ !\frac{\pi^2}{L^2}$. The
%solution can also be written in terms of !the flux function
%$\Psi(r,z)$:
may be written in terms of a poloidal flux function $\Psi(r,z)$
\begin{eqnarray}
&& \Psi=rJ_1\left(j_{11}\frac{r}{R}\right)\cos\frac{\pi z}{L}
\mbox{,}\label{eqn3} \\
&& U_r=-\frac{1}{r}\frac{\ptl \Psi}{\ptl z}\mbox{,}\quad
U_z=\frac{1}{r}\frac{\ptl \Psi}{\ptl r}\mbox{,}\quad
U_{\phi}=\frac{\lambda_B \Psi}{r} \label{eqn4}
\end{eqnarray}
where $J_1(x)$ is a first order Bessel function in its standard notation, and $j_{11}$
is the first root of $J_1(x)=0$,
$\lambda_B^2=j_{11}^2/R^2+ \pi^2/L^2$. The velocity
components $U_r$, $U_z$, and $U_\phi$ have been normalized to a
characteristic velocity $U_0$.

The kinematic dynamo problem for Beltrami flows satisfying
Eqs.~(\ref{eqn3}) and (\ref{eqn4}) is implemented numerically as follows. The flow
field is taken as given and the back reaction of the growing magnetic field on
the flow is neglected. This approximation is justified for the
initial stage of the exponential dynamo growth, when the magnetic
field is weak and does not influence the flow.
Instead of solving the induction equation Eq.~(\ref{induction:1})
for ${\bf B}$ directly, potentials ${\bf A}$ and
$\varphi$ are introduced. Using the gauge condition $\varphi-{\bf U}\cdot{\bf
A}+(\eta /\mu_0)\nabla\cdot{\bf A}=0$~\cite{Finn:2001}, one can derive the following equation of evolution for the vector potential ${\bf A}$,
\begin{equation}
\frac{\ptl A^i}{\ptl t}=-A_k \frac{\ptl U^k}{\ptl x^i}-
U^k\frac{\ptl A^i} {\ptl x^k}+\frac{\eta}{\mu_0}\frac{\ptl^2 A^i}{\ptl x^k \ptl
x_k}\label{eqn6}\mbox{,}
\end{equation}
where the resistivity $\eta$ is assumed to be constant
throughout the cylinder and the coordinate notations refer to a
Cartesian coordinate system $x^i$. A 3D kinematic dynamo
code is used to solve Eq.~(\ref{eqn6}). Then, the magnetic field can
be obtained at any time by taking the curl of ${\bf A}$. The code
is written for cylindrical coordinates, and it uses an explicit scheme
with central spatial differencing in the advection term and
standard nine points stencil for the diffusion term. Since the
conductivity of metallic walls is much higher than the
conductivity of the plasma, a perfectly conducting boundary is a
good approximation. All boundaries of the cylinder
($r=R$, $z=-L/2$, and $z=L/2$) are assumed to be perfect conductors. Then, the
boundary conditions for ${\bf A}$ at perfectly conducting
boundaries compatible with the gauge can be chosen as follows: both the
components of ${\bf A}$ parallel to the boundary and the
divergence of ${\bf A}$ at the boundaries are zero. This gives
three boundary conditions for three components of vector potential. Eq.~(\ref{eqn6}) has a unique solution. A detailed
description of the code, of the gauge choices, and of the
implementation of the boundary conditions can be found in
Ref.~\cite{pariev01}.

In the case of axisymmetric flow, the nonaxisymmetric modes of the
magnetic field, which are proportional to $\exp (in\phi-i\omega t)$  with different
azimuthal wavenumbers $n$ (also known as toroidal mode numbers), are
decoupled from each other and are eigenmodes 
of Eq.~(\ref{induction:1}) or Eq.~(\ref{eqn6}). The 3D
kinematic code picks up the fastest growing mode of the dynamo, which turns out
to be the $n=1$ mode. Since the structure of each eignemode is two
dimensional in $(r,z)$ coordinates, a 2D code involving only the azimuthal component of the magnetic field saves much time
for simulations with varying boundary parameters.
Such a 2D code for the vector potential evolution has also been written and
used to calculate the dependencies of the growth rates of the $n=1$ mode on the boundary dimensions. The growth
rates and the structure of the $n=1$ modes obtained using the 3D dode agree remarkably well with that using the 2D code. The modes have an oscillatory
nature, i.e. the oscillation frequency 
$\omega=\omega^{\prime}+i\gamma (U_{max}/L)$ has
real and imaginary parts, where both the rotation frequency $\omega^{\prime}$ and
the dimensionless growth rate $\gamma$ are real. $U_{max}/L$ is about
100~kHz. $U_{max}$ is the maximum absolute velocity
inside the cylinder. Exponentially growing (or decaying) 
magnetic fields also rotate with certain frequency, which comparable to
the frequency of the fluid rotation, $\omega^{\prime} \sim 100$~kHz.

Dependence of the growth rate $\gamma$ on the
magnetic Reynolds number $Re_m$ and on the aspect
ratio $R/L$ of the cylinder is explored. 
Fig.~3a shows the dependence of $\gamma$ on $Re_m$ for a fixed aspect ratio $R/L=1$. Negative values of $\gamma$ correspond to decaying magnetic
field. For $Re_m$ smaller than the threshold value of 210,
the dynamo will not grow. With increasing $Re_m$, the
growth rate first increases up to the maximum value $\sim
0.022$ (a growth time of
0.45 msec) at $Re_m=500$, and it subsequently decreases to
smaller values but remains positive. Such behavior is typical for slow dynamos
(for example, see Ref.~\cite{Parker:1992}). Thus, these laminar plasma
dynamos are slow dynamos, for which $\gamma$ asymptotically goes to zero
for $Re_m\to\infty$. The laminar and regular (as opposed to chaotic) fluid motion stretches the magnetic field linearly in time. Exponential growth is only possible due to small diffusivity of the magnetic field, which recreates the component of the magnetic field perpendicular to the flow. When the diffusion of the magnetic field becomes very small ($R_m \rightarrow \infty$), the growth rate $\gamma$ also decreases to 0. Fig.~3b shows the
dependence of $\gamma$ on the aspect ratio $R/L$ for a fixed $Re_m=500$. Dynamo
does not exist for very long or very short cylinders. It is
necessary to have comparable $R$ and $L$ for efficient
excitation of the dynamo. In particular, at an experimentally
feasible value of $Re_m=500$, the maximum growth rate
$\gamma=0.055$ is achieved at $R/L=0.6$. These results suggest that a
somewhat elongated cylindrical vessel will be the best for the
excitation of dynamo. Pulsed plasma flow of several msec long should be sufficient to excite the laminar dynamos. 

In conclusion, a theoretical study has shown dynamo can be excited to convert plasma kinetic energy into magnetic field energy in a laboratory
environment. These plasma dynamos are laminar
because of the low kinetic Reynolds number. Modest plasma parameters
self-consistently satisfy the conditions needed for the dynamo.  Numerical
calculation yields a threshold magnetic Reynolds number of 210 for
exponential growth of the laminar plasma dynamos in a cylindrical boundary
with Beltrami flows. These results indicate that a new type of laboratory
dynamo experiments is possible.

We thank Drs.~Hui Li and Stirling Colgate for many useful
suggestions to this work. One of us, VIP, would also like to thank Dr. John
Finn for his advice on numerical simulations. This work is supported by 
U.S. DOE~Contract No.~W-7405-ENG-36. VIP acknowledges partial support from DOE grant DE-FG02-00ER54600

%\pagebreak
%{ \centerline{\bf{Figure Captions}}}
%\vspace{1cm}
%Figure 1, Contour plot for equal potential lines and flow field.

\newpage

 \title {\centering \bf Figures}
  \draft
  \input{epsf}
\maketitle

%DRE
\begin{figure}[hb]
\centering
\vspace{0.5 in}
\epsfig{figure=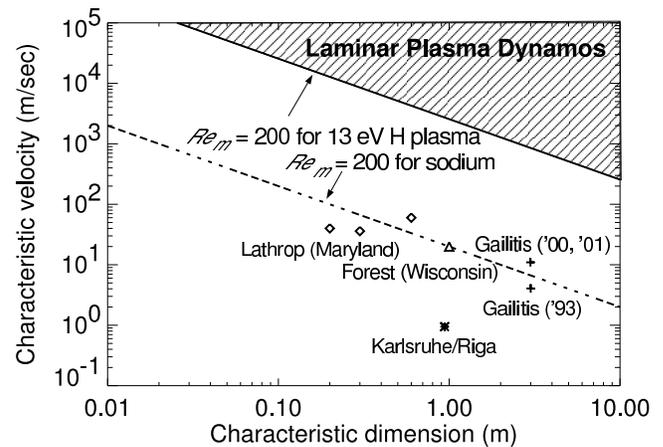,height=2.3 in}
\caption{~Flow velocity vs experimental size for various sodium dynamos and
for laminar plasma dynamos (Shaded region). Specific region of the laminar plasma dynamos may vary with gas species.}
\label{Ilimit.ps}
\end{figure}

\begin{figure}[hb]
\centering
\epsfig{figure=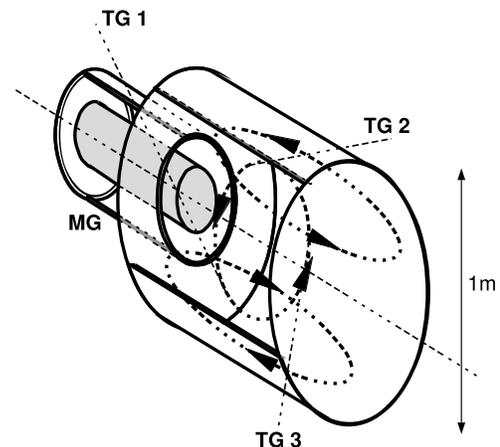,height=2.3 in}
\caption{~A possible laminar plasma dynamo configuration using coaxial plasma guns.}
\vspace{0.5 in}
\label{Ilimitb.ps}
\end{figure}
\newpage

\begin{figure}[hb]
\centering
\epsfig{figure=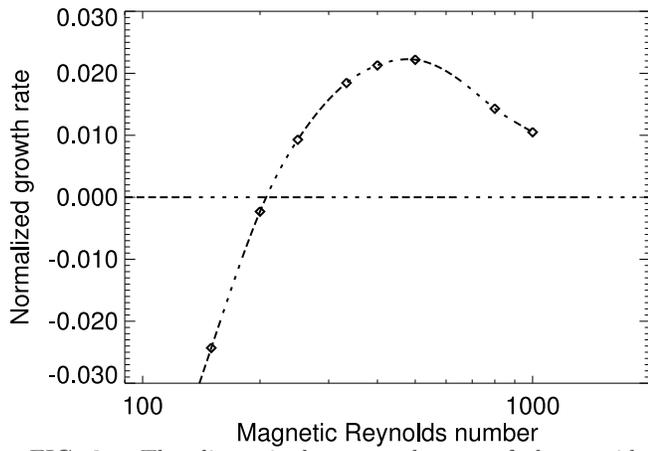,height=2.3 in}
\caption{~The dimensionless growth rate of the toroidal mode
$n=1$ of a laminar plasma dynamo ($\gamma$) as a function of  the magnetic Reynolds number, $Re_m$, for aspect ratio $R/L = 1$.}
\vspace{0.5 in}
\label{Fig3a.ps}
\end{figure}

\begin{figure}[hb]
\centering
\epsfig{figure=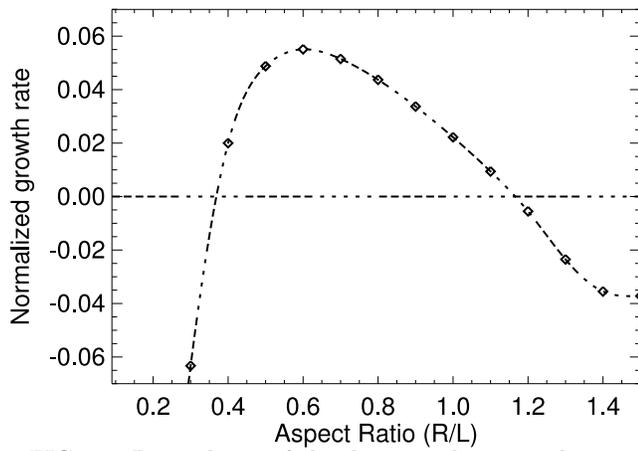,height=2.3 in}
\caption{~Dependence of the dimensionless growth rate of the toroidal mode
$n=1$ of a laminar plasma dynamo ($\gamma$) on the aspect ratio, $R/L$, for
fixed magnetic Reynolds number $Re_m= 500$.
}
\vspace{0.5 in}
\label{Fig3b.ps}
\end{figure}

\end{document}